\documentclass[a4paper,fleqn,usenatbib]{mnras}

\usepackage{mathptmx}
\usepackage{threeparttable}

\usepackage[T1]{fontenc}
\usepackage{ae,aecompl}
\usepackage{deluxetable}
\usepackage{graphicx}	% Including figure files
\usepackage{amsmath}	% Advanced maths commands
\usepackage{amssymb}	% Extra maths symbols
\usepackage{txfonts}

\usepackage{mathrsfs}
\usepackage{amsmath}
\usepackage{bm}
\usepackage{epstopdf}
\usepackage{graphicx}

\def\be{\begin{equation}}
\def\ee{\end{equation}}
\def\bea{\begin{eqnarray}}
\def\eea{\end{eqnarray}}
\def\f{\frac}

\def\rm{\textrm}
\title{The Synchrotron Polarization In Decaying Magnetic Field in Gamma-Ray Bursts }

\author[K. F. Cheng et al.]{
	Cheng K. F.$^{1,4}$,
	Zhao X. H.$^{1,2,3}$\thanks{E-mail:zhaoxh@ynao.ac.cn},
	Bai J. M.$^{1,2,3}$\\
	$^1$Yunnan Observatories, Chinese Academy of Sciences, Kunming, China; \\
	$^2$Center for Astronomical Mega-Science, Chinese Academy of Sciences, Beijing, China \\
	$^3$Key Laboratory for the Structure and Evolution of Celestial Objects, Chinese Academy of Sciences, Kunming, China \\
	$^4$University of the Chinese Academy of Sciences, Yuquan Road 19, Shijingshan Block, 100049, Beijing, people's Republic of China\\}

\date{Accepted XXX. Received YYY; in original form ZZZ}

\begin{document}
\label{firstpage}
\pagerange{\pageref{firstpage}--\pageref{lastpage}}
\maketitle

\begin{abstract}
Polarization can serve as a probe of the radiation mechanism and magnetic field (MF) configuration in gamma-ray bursts (GRBs). In the case of constant MF, the synchrotron polarization in the prompt phase of GRBs has been widely studied. In this paper, we consider the case of the decaying MF. We calculate the time-averaged and instantaneous synchrotron polarizations in a pulse for different viewing angles and for the large-scale decaying MF model, which can explain the so-called Band spectrum. We find that the on-axis time-averaged polarization degree (PD) in the energy band of 50-500 keV for the decaying large-scale MF model ($\sim 0.6$ for typical parameters) is higher than that in the constant MF model ($\sim 0.5$). An interesting result is the instantaneous PD in the off-axis case will experience a turnover, i.e., the PD will evolve from a positive value to a negative one. This suggests the polarization angle (PA) change by an angle of $90^\circ$. Such a result is roughly consistent with the discovery of the PA evolution within a pulse in some bursts, such as GRB 170114A and GRB 160821A. Our result implies at least a part of bursts (off-axis bursts) should have the PA evolution in a pulse.

\end{abstract}

\begin{keywords}
gamma-ray burst: general - polarization - magnetic fields - radiation mechanisms: non-thermal
\end{keywords}

\section{Introduction}
Gamma-ray bursts (GRBs) are the most intense electromagnetic explosions in the universe after the big bang. The prompt GRB emission is widely believed to be produced by internal dissipation while the afterglow comes from the external shock in the classical fireball model. Observations have supported the afterglow is produced by the synchrotron emission of electrons in the external shock. While the radiation mechanism of the prompt emission still remains a mystery to date due to the limited information from the observed spectra. Synchrotron emission from relativistic electrons is one of the most important candidates. However, the observed spectra of GRBs are inconsistent with the synchrotron model with a constant MF (however, see Burgess et al. 2020). The observed GRB spectrum is usually well fitted by the so-called Band function (Band 1993) composed by low and high energy power laws with a smooth connection. The low-energy spectral
index $\alpha$ $(N_{\nu} \propto \nu^{\alpha})$ is $\sim -1$ (Preece et al. 2000; Kaneko et al. 2006), while the theoretical synchrotron spectrum of fast cooling electrons is $-3/2$ (Sari et al.1998). This is the so-called fast cooling problem (Ghisellini et al. 2000).
Hence, some modified synchrotron radiation models have been proposed to solve this problem.
The Klein-Nishina (KN) inverse Compton (IC) cooling can change the fast cooling electron slope, and leads to a more harder spectrum with a slope of $\sim -1$,
which is roughly consistent with observations (e.g., Wang et al. 2009; Nakar et al. 2009). The large-scale or small-scale decaying MF in GRB emission region with the adiabatic cooling and the IC cooling also
lead to a harder synchrotron spectrum than the fast cooling spectrum, which can also explain the Band spectrum (Uhm \& Zhang 2014, hereafter UZ14; Pe'er \& Zhang 2006; Zhao et al. 2014). Geng et al.
(2018, hereafter Geng18) added one case of the combination of the KN effect and the large-scale decaying MF and also gave similar results. However, the validation of the models needs a
further check. Some authors have fit the data with the models (Zhang et al 2016).

The synchrotron emission of GRB would have a high linear polarization if the magnetic field (MF) in the GRB emission region is ordered. In the current internal dissipation mechanisms of prompt emission, such as internal shocks or magnetic reconnection (e.g., Rees \& M\'{e}sz\'{a}ros 1994; Zhang \& Yan 2011), electrons in GRB emission region are accelerated into relativistic speed with a non-thermal power-law distribution, which can be described by $dN'_e/d\gamma'_e \propto \gamma^{'-p}_{e}$, where $\gamma'_e$ is the electron energy and $p$ is electron index. Note that we prime all the quantities in the GRB jet comoving frame. Hence, the intrinsic polarization degree (PD) of synchrotron emission is calculated as $\Pi_{\rm{syn}} =(p+1)/(p+7/3)$ (Rybicki \& Lightman 1979). For the typical value of $p=2.8$, we get $\Pi_{\rm{syn}} \simeq 74\%$.
If the emission of GRBs is from both non-thermal and thermal electrons, where the thermal electrons with a Maxwellian distribution can be produced by the second-order Fermi acceleration mechanism (Schlickeiser 1985; Stawarz \& Petrosian 2008; Giannios \& Spitkovsky 2009), the mixture of these two electrons components will lead to a significant change of the synchrotron polarization (Mao \& Wang 2018).

The synchrotron polarization of GRB essentially comes from the asymmetry of the MF. The MF is believed to have two origins. One is shock. The Weibel instability in shock will generate a small-scale random field (e.g., Gruzinov \& Waxman 1999; Medvedev \& Loeb 1999), where the MF coherent scale is much smaller than the GRB emission region. The MF orientation in each coherent region is random, so if the visible region of the jet is symmetric, the net polarization is nearly 0. In the afterglow phase, the problem is widely investigated. The net polarization can be high when the jet edge is observed due to the jet deceleration and the symmetry of the visible region is broken (e.g., Sari 1999; Granot \& K$\ddot{o}$nigl 2003).   Another origin is the large-scale order MF (e.g., Spruit et al. 2001; Lazzati 2006), which comes from the central objects of GRB. The MF have several kinds of configuration, such as poloidal and toroidal fields. The toroidal field decay more slowly with the expanding of the emission region, compared with the poloidal field and thus should be dominated in the MF model. The PD in this MF will have a higher level than that in the small-scale random MF. In addition, jet structure of GRB (e.g., top-hat jet; structured jet; two-component jet) has an important effect on the polarization level and its evolution (e.g., Rossi et al. 2004; Gill et al. 2020). Therefore, polarization measurement of GRBs is an effective tool to study the radiation mechanism of GRB, MF configuration and jet structure.

Recently, increasing polarization detections were reported, especially in the prompt phase (see the compilation of Gill et al. 2020). For instance, Gamma-Ray Burst polarimeter (GAP) detected GRB 100826A with an average PD of $27 \pm 11 \%$ with $99.4\%$ $(2.9\sigma)$ confidence level (Yonetoku et al. 2011).
GRB polarimeter POLAR onboard China's Tiangong-2 spacelab has detected a sample of GRBs
with a relatively low level of PD, compared with other polarimeters, such as GAP, AstroSat-CZTI and INTEGRAL-IBIS, and found the evolution of polarization angle (PA) in a single pulse for GRB 170114A (Zhang et al. 2019a). The averaged PD over the whole sample is $\sim10\%$. Such a relatively low level of PD is also partly attributed to the intrapulse evolution of PA (Zhang et al. 2019a). Significant interest is aroused by the polarization observations (e.g., Burgess et al. 2019; Chattopadhyay et al. 2019; Lan \& Dai 2020; Gill et al. 2020), though most of these detections possess a confidence level of $<5\sigma$ (however, see Sharma et al. 2019).

The polarization of the "standard" synchrotron model (with constant MF) in the prompt phase has been widely studied (e.g., Nakar et al. 2003; Toma et al. 2009, hereafter Toma09). 
However, as mentioned above, the MF in GRB emission region could be a decaying field. The synchrotron polarization in this kind of field is worth to investigate. In this paper, we calculate the instantaneous and time-averaged
synchrotron polarization with different MF models. Our motivation is trying to identify the MF models , to verify the GRB radiation mechanism with the observed polarization data and to make preparation in theory for the high-precision PD data. The structure of this paper is as follows.
The assumed decaying MF model and the calculation of electron distribution evolution are presented in section 2 and the resulting spectra are also given. In section 3, we calculate the instantaneous and time-averaged polarization with large-scale MF models. We summary our results and discuss the prospect for future observations in section 4.

\section{electron distribution and the spectra with decaying MF}
Consider that GRB is produced by synchrotron radiation of high energy electrons. The electrons can be generated by some processes, such as magnetic
reconnection (Zhang \& Yan 2011) and internal shocks (e.g., Rees \& M\'{e}sz\'{a}ros 1994). Initially the electrons are believed to be a power-law distribution while the distribution is modified by several cooling mechanisms including synchrotron (SYN), adiabatic (ADI), and synchrotron self-Compton (SSC) coolings. The MF models, such as the decaying fields, will significantly affect the cooling rates of the electrons and lead to different electron distributions and spectra.

\subsection{MF Model}
The MF configuration in the emission region of GRBs is unknown.
A possibility is the MF is ordered on large scales, such as the toroidal MF, which comes from the GRB central engine. The field couples with the GRB jet material and is launched along
with the jet. Due to the expanding of the jet, the MF strength in the jet comoving frame will decay with the jet radius. This will produce a harder spectrum than the fast cooling spectrum, which can explain the GRB
spectra (UZ14). Consider a GRB comes from a thin shell moving with a bulk Lorentz factor of $\Gamma$. The MF strength can be described by (Spruit et al. 2001; UZ14)
\begin{equation}
B'=B'_0 (\frac{R}{R_0})^{-a},
\end{equation}
where $R=R_s+\beta c\Gamma t'$ is the jet radius, $R_0$ is the radius where the MF begins to decay, $R_s$ is the radius where the GRB emission starts, $B'_0$ is the initial MF strength and $a$ is the MF
decaying index. $a=0$ corresponds to the constant MF case.

\subsection{Electron distribution and spectra}
In the GRB emission region, high-energy electrons mainly cool down by SYN, ADI and IC mechanisms. For an electron with energy of $\gamma'_e$,
the synchrotron cooling rate is given by (Rybicki \& Lightman 1979)
\begin{equation}
\dot{\gamma}'_{e,\rm{syn}}=-\frac{\sigma_{T} B^{'2} \gamma_{e}^{'2}}{6 \pi m_{e} c}.
\end{equation}
The ADI cooling rate can be described as (e.g., Tavecchio et al. 2003; Uhm et al.2012)
\begin{equation}
\dot{\gamma}'_{e,\rm{adi}}=-\frac{2}{3} \frac{\gamma'_{e}}{R} \frac{dR}{dt'}=-\frac{2}{3} \frac{\beta c \Gamma \gamma'_{e}}{R}.
\end{equation}

The IC cooling rate is (Jones 1968; B\"{o}ttcher et al. 1997; Finke et al. 2008; Yan et al. 2016)

\begin{equation}
\dot{\gamma}'_{e,SSC}=-\frac{3 \sigma_T m_e c^3}{8 h^2 }
\int_{0}^{\infty} \frac{u'(\nu')}{\nu'^2} {\cal G}(\gamma'_{e},\nu')d\nu'
\end{equation}
where ${\cal G}(\gamma'_{e},\nu')$ are given by
${\cal G}(\gamma'_{e},\nu')=\displaystyle\frac{8}{3}\frac{E(1+5E)}{(1+4E)^2}-\frac{4E}{1+4E}(\frac{2}{3}+\frac{1}{2E}+\frac{1}{8E^2})+\rm{ln}(1+4E)[1+\frac{3}{E}+\frac{3}{4}\frac{1}{E^2} 
\displaystyle+\frac{\rm{ln}(1+4E)}{2E}-\frac{\rm{ln}(4E)}{E}] -\frac{5}{2}\frac{1}{E}+\frac{1}{E}\sum_{n=1}^{\infty}
\frac{(1+4E)^{-n}}{n^2}-\frac{\pi^2}{6E}-2 $ for $E>1/4$, and ${\cal G}(\gamma'_{e},\nu')=\displaystyle E^2 (\f{32}{9}-\f{112}{5}E+\f{3136}{25}E^2)$ otherwise. $E=\gamma'_{e}h\nu'/m_{e}c^2$ and $h$ is the Planck constant. Here we adopt a differet method from that of Geng18 in
calculating the IC cooling rate. The method should be more efficient in numerical implementation because the dimension of integral is less. $u'(\nu')$ is the comoving photon energy density per unit frequency, and is given by
\begin{equation}.
\begin{cases}
u'(\nu')=4 \pi I'(\nu')/c \\
I'(\nu')=j'(\nu') \Delta R' \\
\displaystyle j'(\nu')=\frac{1}{4\pi} \int_{\gamma'_{\rm m}}^{\gamma'_{\rm{max}}}P'(\gamma'_e,\nu')\frac{dN'_e}{d\gamma'_e}\f{1}{V'}d\gamma'_e
\end{cases}
\end{equation}
where $\gamma'_{e,\rm m}$ and $\gamma'_{e,\rm{max}}$ are the minimum and maximum electron Lorentz factors, respectively
and $P'(\gamma'_e,\nu')$ is the spectral power (erg s$^{-1}$ Hz$^{-1}$) of the synchrotron emission in the comoving frame (Rybicki \& Lightman 1979)
\begin{equation}
P'(\gamma'_e,\nu')=\frac{\sqrt{3}q_{e}^3 B' \sin\alpha'}{m_e c^2}[x\int_{x}^{\infty} K_{5/3}(\xi)d\xi].
\end{equation}
Here $x=\nu'/\nu'_{c}$, $\nu'_c=\frac{3q_{e}B'\sin\alpha'}{4\pi m_e c}\gamma_{e}^{'2}$, and $K_{5/3}(\xi)$ is the Bessel function,
$\alpha'$ is the pitch angle, i.e., the angle between the magnetic field and the electron velocity. $\Delta R' \thickapprox R/\Gamma $ is the comoving width of the shell. $I'(\nu')$ and $j'(\nu')$ are the specific intensity and emission coefficient (Rybicki \& Lightman 1979), respectively. $V'=4\pi R^2\Delta R'$ is the volume of the shell. The instantaneous electron distribution $dN'_e/d\gamma'_e$ can be obtained by solving the continuity equation of the electrons in energy
space (e.g., Longair 2011; Zhao et al. 2014; UZ14; Geng18),
\begin{equation} \label{continu_equ}
\frac{\partial}{\partial t'}\left(\frac{dN'_e}{d\gamma'_e}\right) + \frac{\partial}{\partial \gamma'_e}
\left[\dot{\gamma}'_{e,tot}\left(\frac{dN'_e}{d\gamma'_e}\right)\right] = {Q'}(\gamma'_e)
\end{equation}
where $\dot{\gamma}'_{e,\rm{tot}}$ is the total cooling rate of the electrons, i.e., $\dot{\gamma}'_{e,\rm{tot}}=\dot{\gamma}'_{e,\rm{syn}}+\dot{\gamma}'_{e,\rm{adi}}+\dot{\gamma}'_{e,\rm{SSC}}
$. $Q'(\gamma'_e)=Q'_{0}\gamma_{e}^{\prime-p} $ (for $\gamma'_{m}<\gamma'_{e}<\gamma'_{\rm{max}} $) is
the electron injection term, which is assumed to be a power-law form.

The observed flux density is $F_\nu\simeq I'(\nu')\delta^3\Delta\Omega/(1+z)^3$, where $\Delta\Omega=\pi (R/\Gamma)^2/D_A^2=\pi R^2(1+z)^4/\Gamma^2D_L^2$ is the solid angle of the visible region seen from the observer, the comoving frequency is $\nu'=(1+z)\nu/\delta$, z is the redshift and $D_L$ is the luminosity distance. Taking the approximation of Doppler factor $\delta$ in the visible angular range of $1/\Gamma$ as $\delta \approx \Gamma$ and using equation (6), we obtain
\bea
F_\nu\approx\f{\Gamma(1+z)}{2\pi D_L^2}\int^{\gamma'_{\rm{max}}}_{\gamma'_m} P'(\gamma'_e,\nu')\f{dN'_e}{d\gamma'_e}d\gamma'_e.
\eea
With this equation, we can calculate the flux density spectra.

\subsection{models}
Similar to what was done in Geng18, we consider three kinds of models with different cooling mechanisms, namely, M1 (only synchrotron cooling with the constant MF or $a=0$), M2 ( synchrotron and adiabatic coolings with $a=0.0, 1.0, 1.2, 1.5$, corresponding to the model of UZ14) and M3 (synchrotron, adiabatic and SSC coolings with $a=0.0, 1.0, 1.2, 1.5$, corresponding to M3 and M4 of Geng18).
The detail model parameters used in our calculations are listed in Table 1- 3. The adopted parameters, such as $B'_0$ and $\gamma'_m$, are based on the constraint of observed spectral peak energy $E_{\rm{peak}}=1/(1+z)\cdot 3hq_eB'_0\Gamma\gamma_m^{\prime 2}/4\pi m_e c\sim 500 \rm{keV}$, where the on-axis case is assumed. The cosmological parameters ($\Omega_m=0.27$, $\Omega_\lambda=0.73$, $H_0$=71 km s$^-1$ Mpc$^{-1}$) are used, and the redshift is set to be $z=1$.

We solve the continuity equation of electrons in the energy space (Eq. \ref{continu_equ}). The fully implicit difference scheme proposed by Chang \& Cooper (1970)
and Chiaberge \& Ghisellini (1999) is adopted. To test our code, we first use the same parameters as those in UZ14 and Geng18, and compare the
resulting electron distributions and spectra. With the parameters listed in Table 1, we find our results are consistent with theirs (see Fig. 1 and 2). It needs to be noted that the initial MF strength in UZ14, where the GRB starts, is actually 300 G for
$a=1$, not 30 G,  because the MF begins to decay at $R_0=10^{14}$ cm while the GRB emission starts from $R_s=10^{15}$ cm in their model.
In Fig. 3 and 4, like Geng18, we consider the case in which the two radii are equal. One can find the decaying MF indeed
generates a harder spectrum in low-energy bands, similar to the Band spectrum. With the electron distribution we will calculate the PD.

\section{synchrotron polarization in the large-scale decaying MF model}
\subsection{Formulation}
In this section, we describe the method of calculating the polarization of synchrotron emission which with a large-scale order MF.
A relativistic electron in the large-scale order MF will generate strong polarized synchrotron radiation. The spectral power of the electron
can be divided into two different polarization states, i.e., (Rybicki \& Lightman 1979)
\begin{equation}
\begin{cases}
P'_{\perp}(\nu')=\frac{\pi \sqrt{3} q_{e}^{2} \nu'_{L}\sin\alpha'}{c}[F(x)+G(x)] \\
P'_{\parallel}(\nu')=\frac{\pi \sqrt{3} q_{e}^{2} \nu'_{L}\sin\alpha'}{c}[F(x)-G(x)]
\end{cases}
\end{equation}
where
\begin{equation}
\begin{cases}
F(x)=x\int_{x}^{\infty}K_{5/3}(\xi)d\xi \\
G(x)=xK_{2/3}(x)
\end{cases}
\end{equation}
and $\nu'_{L}=q_{e}B/2\pi m_{e}c$ is the gyro-frequency.
The synchrotron polarization from the electron can be calculated by (Rybicki \& Lightman 1979)
\begin{equation}
\Pi'_{0}=\frac{P'_{\perp}(\nu')-P'_{\parallel}(\nu')}{P'_{\perp}(\nu')+P'_{\parallel}(\nu')}=\frac{G(x)}{F(x)}
\end{equation}
%
%The synchrotron radiation of GRBs comes from a group of injected electrons,
For an electron population with a distribution of $dN'_e/d\gamma'_e$, the total polarization degree is
\begin{equation}
\Pi_{\rm{syn}}(\nu')=\frac{\int_{\gamma'_{\rm{m}}}^{\gamma'_{\rm{max}}}G(x)\frac{dN'_{e}}{d\gamma'_e}  d\gamma'_{e}}
{\int_{\gamma'_{\rm{m}}}^{\gamma'_{\rm{max}}} F(x) \frac{dN'_{e}}{d\gamma'_e}  d\gamma'_{e} }.
\end{equation}
Consider the GRB jet has an half opening angle of $\theta_j$ and assume the jet is dominated by the
toroidal magnetic field. The synchrotron polarization for a given observed frequency $\nu$ from the burst can be derived by calculating the Stokes parameters, i.e.,
%$Q_{\nu_{\rm{obs}}}=I_{\nu_{\rm{obs}}}  \Pi_{\rm{syn}} \cos(2\chi) $ and $U_{\nu_{\rm{obs}}}=I_{\nu_{\rm{obs}}}  \Pi_{\rm{syn}} \sin(2\chi) $, or
%
\begin{equation}
\begin{Bmatrix}
Q_{\nu}  \\
U_{\nu}
\end{Bmatrix}
=\frac{1+z}{4 \pi d_{L}^{2}}\int d\phi \int  \delta^{3} P'(\nu') \Pi_{\rm{syn}}(\nu')
\begin{Bmatrix}
\cos(2\chi) \\
\sin(2\chi)
\end{Bmatrix}
d(\cos \theta),
\end{equation}
where $\chi$ is the polarization position angle in the observer frame. $\nu'=(1+z)\nu/\delta$ is the comoving frequency and $\delta=1/[\Gamma(1-\beta
\cos \theta)]$ are the Doppler factor.
For the given observed wavebands [$ \nu_{1} , \nu_{2} $], the degree of linear polarization of the jet emission is given by
\begin{equation}\label{pi}
\begin{aligned}
\Pi =\f{Q}{I}&=\int_{\nu_1}^{\nu_2}d\nu \int_{0}^{(1+q)^{2}y_{j}} f(y) dy \int_{-\Delta \phi(y)}^{\Delta \phi(y)} d\phi  \int_{\gamma'_{\rm{m}}}^{\gamma'_{\rm{max}}} G(x)\frac{dN_{e}}{d\gamma'_e}(t') \\
&\times B(t') \sin \alpha' \cos (2\chi ) d\gamma'_{e}  [\int_{\nu_1}^{\nu_2}d\nu \int_{0}^{(1+q)^{2}y_{j}} f(y) dy\\
&\times \int_{-\Delta \phi(y)}^{\Delta \phi(y)} d\phi  \int_{\gamma'_{\rm{m}}}^{\gamma'_{\rm{max}}} F(x)\frac{dN_{e}}{d\gamma'_e} (t') B(t')  \sin \alpha'  d\gamma'_{e} ]^{-1}
\end{aligned}
\end{equation}
where $t'= t_{\rm{obs}}\delta/(1+z)$ is the jet comoving time and $t_{\rm{obs}}$ is the observed time. Note that $U=0$.
Following Toma09, here we define several variables, $y \equiv (\Gamma \theta)^{2}$,  $y_{j} \equiv (\Gamma
\theta_{j})^{2}$, and $q \equiv \theta_{v} / \theta_{j}$, where  $\theta_{v}$ is the viewing angle.
$f$ is a function of $y$. During an observed GRB pulse, the polarization degree varies with time. We thus calculate the instantaneous and time-integrated polarizations over the pulse. The observed pulse can be generated when a thin relativistic shell emits photons within a radius range of $R_s$ to $R_{\rm{off}}$, i.e., the emission starts at $R_s$ and turns off at $R_\rm{off}$. Note that the turning off of the emission does not mean the observed emission sharply stops, because the higher latitude emission arrives at the observer at later time. Take 
the time when a testing photon emitted from the origin (the center of the GRB central engine) arrives at the observer as the observed time zero point. The starting times of the GRB
pulse are $t_{\rm{obs0}}=R_{s}(1+z)[1-\beta]/(\beta c)$ and $t_{\rm{obs0}}=R_{s}(1+z)[1-\beta \cos(\theta_{v}-\theta_{j})]/(\beta c)$ for the on-axis and off-axis cases,
respectively. The turn-off times of the GRB pulse are defined as $t_{\rm{off}}=R_{\rm{off}}(1+z)[1-\beta]/(\beta c)$ and $t_{\rm{off}}=R_{\rm{off}}(1+z)[1-\beta \cos(\theta_{v}-\theta_{j})]/(\beta c)$ for the on-axis and off-axis cases, respectively, where $R_{\rm{off}}$ is set to be $R_{\rm{off}}=10R_{\rm{s}}$. For the time-integrated polarization, $f(y)=(1+y)^{-2}$ while for the instantaneous polarization, $f(y)=(1+y)^{-3}$ (Nakar \& Piran 2003).  Other variables can be described by (Toma09)
\begin{equation}
\sin \alpha' = \left[ \left(\frac{1-y}{1+y}\right)^{2} + \frac{4y}{(1+y)^{2}} \frac{(s-\cos \phi)^{2}}{(1+s^{2}-2s \cos \phi)} \right]^{1/2}
\end{equation}
\begin{equation}
\chi = \phi + \rm{arctan} \left( \frac{(1-y)}{(1+y)} \frac{\sin \phi}{(a- \cos \phi)}\right)
\end{equation}
\begin{equation}
\Delta \phi(y) =
\begin{cases}
0,    \qquad \qquad \qquad \qquad \rm{for} \; q>1  \; \rm{and}  \;  y<(1-q)^2 y_{j}     \\
\pi,  \qquad \qquad \qquad \qquad \rm{for} \; q<1  \; \rm{and}  \;  y<(1-q)^2 y_{j}   \\
\cos^{-1} \left[\frac{(q^{2}-1)y_{j}+y}{2q \sqrt{y_{j}y}}\right]   \qquad \qquad \qquad \qquad \quad \rm{otherwise}.
\end{cases}
\end{equation}
where $s=\theta/\theta_{v}$.

\subsection{instantaneous polarization and light curves}
Fig. 5 shows the instantaneous polarization evolution and the corresponding normalized (normalized to the peak flux) light curves for the constant MF model of M1R14$\Gamma$300(0.0) (see Table 2 for detailed parameters) and for different viewing angles. Here we take the jet opening angle as a typical value of $0.1$ and all the PDs are calculated in the observed energy band of 50-500 keV, which is roughly the energy range of the current GRB polarimeters such as POLAR. 
Fig. 6 and 7 are for models M2 and M3 (constant and decaying MF models with different initial radii), respectively. 
One can find generally the PDs in various models for both the on-axis and off-axis cases have similar behaviors, i.e., they first decrease steeply at the beginning, then approximately remain a constant (plateau) before the turn-off time $t_{\rm{off}}$ and decrease rapidly after $t_{\rm{off}}$, except that in Fig. 6 and 7, there is a rising at the very beginning. It is also noted that the PD evolution is nearly independent of the initial radius and that the off-axis PD decreases with the increase of the viewing angle.

The PD curves are different in details (see Fig. 5, 6 and 7). The off-axis PD is higher than the on-axis PD at the beginning of the pulse. This can be
understood as follows. In the very early phase, the polarization is approximately the local physical polarization
since only a very tiny angular region is observed and the MF can be seen as parallel lines. The intrinsic energy band contributing to the given observed band would be higher in the off-axis case than in the on-axis case
due to the Doppler effect. So at an early time, the off-axis PD can be from the -p/2 spectral segment of the synchrotron emission and
thus is as large as $\sim0.78$ (p=2.8), while the on-axis PD is from the combination of
1/3 and -1/2 spectral segments and is in the range of $0.5-0.7$. If the electron cooling is relatively slow (e.g., for weaker MF in 
Fig. 6 and 7), the injected minimum-energy electrons has not cooled yet at enough early time. The PD will evolve from 1/3 spectral
segment dominated to -1/2 and then to -p/2 dominated. Bear in mind that the electrons contributing to the given energy band
would move toward higher energies due to the MF decay. So we find there is a rising phase in Fig. 6 and 7 With the electron
cooling, both the electron distribution changes and the geometric effect that larger angular region with curved MF enters the line
of sight will work and thus the PD begins to decrease.

For each model, there is a plateau before $t_{\rm{off}}$ in the PD curves. The PDs in the plateau are $\sim 0.6$ in the on-axis case, in contrast to $\sim0.2$ in the off-axis case. For the on-axis case ($q=0.1$) the PD plateaus of the decaying MF models are a little higher than those of the constant MF models, while for the off-axis case the results are the opposite. The formation mechanism of the plateau is due to the fact that before $t_{\rm{off}}$, the dominated emission is from the same angular areas around $\theta \sim 0$ for the on-axis case or around $\theta\sim \theta_v-\theta_j$ for the off-axis case but from different radii. Thus the same MF configuration in the areas produces the same PDs. After $t_{\rm{off}}$, the emission of the pulse is from the high latitudes ($\theta>0$ for on-axis case or $\theta> \theta_v-\theta_j$ for off-axis case). The PDs are partly cancelled out due to the increasing curvature of the MF (see Fig. 8) and thus produces a rapidly decreasing net PD. 

It is noteworthy that the off-axis PD changes from positive to negative shortly after $t_{\rm{off}}$, suggesting the PA
varies by an angle of 90$^\circ$, and turns to rise at more late time. This can be understood from Fig. 8. At the beginning in the off-axis case, the MF is
horizontal direction dominated (area 1) and thus the net polarization is in the vertical direction, while at a late time, the MF turns to the vertical direction dominated (area 2) and thus the net
polarization is in the horizontal direction. At a later time the horizontal MF is again dominated (area 3) so that the polarization is in the vertical direction. Actually, this behavior of the PA also happens in the on-axis case, if only the line of sight (LOS) is not right on the axis of the jet symmetry. But for the on-axis case, the flux from the nearby region of the LOS is much larger than that from the high latitude region, rendering the polarization detection in the high latitude not easy. 

Regarding the light curves in Fig. 5, 6 and 7, we can find for the constant MF, their peaks are consistent with $t_{\rm{off}}$, while for the decaying MF, their peaks appear much earlier than $t_{\rm{off}}$ and depend on the MF decaying indexes. For on-axis cases, the larger MF decaying indexes, the earlier peak times, which are consistent with the results of Uhm \& Zhang (2016) or Uhm et al. (2018). The light curve before $t_{\rm{off}}$ is determined by three factors: the spectral power of electrons, the number of the emitting electrons contributing to the observed band and the spectral evolution in the given observed energy band. The number of the emitting electrons increases with time due to enhancing emission from larger angles from the LOS. This factor will lead to a rise of light curves. The spectral power of electrons declines from the beginning of the pulse due to the MF decay and lead to a decline of light curves. At the beginning, the spectral peak ($E_p$) is in the observed band (50-500 keV) and roughly lead to an unchanged flux. However, with the decrease of $E_p$ and its crossing the observed band ($<50$ keV), the light curve will decline because the observed band enters the high energy band in which the flux steeply declines with energy. The combined effect of the three factors gives rise to the light curves in the figures. The different light curve peaks in the decaying MF case mainly arise from the spectral peak energy $E_p$ crosses the observed energy band. We verified this by adopting a wider spectral range such as 1-500 keV and finding the peak appears later.

For the off-axis cases, $E_p$ have crossed the observed band ($<50$ keV) since the beginning of light curves and thus the light curves arrive at their peaks very early. Combining the light curves and the PD curves, we can find that: the early steep decline of PDs corresponds to the rise of light curves; the plateaus corresponds to the time between the light curve peak and $t_{\rm{off}}$; the steep decline following the plateaus corresponds to the light curve decline due to the high-latitude emission (curvature effect), which suggests the PD evolution after $t_{\rm{off}}$ is due to the curvature effect.

\subsection{time-averaged polarization}
Fig. 9 and Fig. 10 show the time-averaged PD within a pulse as a function of the viewing angle (normalized to the jet opening angle) for model M2 and M3. The PD profiles are similar to those of Toma09 or Gill et al.
(2020). But the PD are systematically higher than those of Toma09. The difference is due to the fact that the decaying
MF produces spectrum with gradually changing from $-p/2$ to $\sim0$ toward lower energies ($F_\nu$ spectrum), while the slopes of the low and high energy bands is set as  fixed values in Toma09. One can find the time-averaged on-axis PDs are $\sim0.6$ and $\sim0.5$ for the decaying and constant MF cases, respectively, while the off-axis PDs in both cases sharply decrease down to $\sim0.1$. The on-axis PD in the decaying MF case is significantly higher than that in the constant MF case. This is due to the fact that the comoving energy band contributing to a given energy band is different for the two cases. For the constant MF case, the comoving energy band contributing to the given observed energy band of $50-500$ keV is mainly in the spectral segment of $-1/2$, while for the decaying MF case, the dominating contribution is from the combination of $\sim0$ and $-p/2$ with increasing contribution of $-p/2$ with time due to the MF decay.

\section{summary and discussion}
Recently, more and more GRB polarizations in the prompt phase are measured and the precision is also increasing (e.g., Yonetoku et al. 2012; Chattopadhyay et al. 2019; Zhang et al. 2019a). This will provide another crucial
tool, in addition to the spectrum, to diagnose the GRB radiation mechanism and the MF configuration. In this paper, we calculate the linear polarization of the emission from a GRB jet with constant and decaying large-scale ordered MFs based on the synchrotron mechanism. The following characteristics are found with different MF models:

1. The time-averaged on-axis and off-axis polarizations in a pulse in the decaying large-scale MF case are PD$\sim0.6$ and PD$\sim0.1$ for plausible GRB parameters, respectively. In contrast, they are $\sim0.5$ and $\sim0.1$ in the constant large-scale MF case.

2. The instantaneous on-axis PD in a pulse has a rise at the very beginning time and then decays, while the off-axis PD decays from the beginning, and then both on-axis and off-axis PDs enter a plateau phase, followed by a steep decay again.

3. Before the turn-off time $t_{\rm{off}}$ of a pulse, the instantaneous on-axis PD in the plateau phase is $\sim0.6$, and then decreases down to $\sim0.1$ after $t_{\rm{off}}$.

4. The instantaneous off-axis PD in the plateau phase is $\sim0.2$ before $t_{\rm{off}}$ and then declines down to even less than $0$ after $t_{\rm{off}}$. This suggests that the PA of the linear polarization undergoes a change of 90$^\circ$.

The PDs are all calculated in the 50-500 keV band. The detection of the time-averaged polarization is more promising than that of the instantaneous polarization due to the low photon statistics in a typical
burst. Given typical observed spectral parameters, the expected PD would be $\sim0.6$ for a large burst sample if the MF is the large-scale decaying field, while it
would be $\sim0.5$ if the MF is the large-scale constant field. This may be as a diagnose of the MF properties. But the difference is small and distinguishing
the two models requires higher precise detection of PD. For a specific burst, we should combine the spectra fitting and PD to determine the MF properties. Some detailed fitting of spectra with the decaying large-scale MF model have been done (e.g., Zhang et al. 2016). In the future, if the polarization detection with high precision is obtained, it will also serve as a test of the model.

According to the results of Zhang et al. (2019a), the PA of GRB 170114A detected by POLAR changed roughly by an angle of $\sim90^\circ$ in two time segments in a
single pulse. The time-integrated PD over the entire pulse is in a low level $\sim 4\%$, while the PD in either segment is larger than 10\%. The PD of GRB 160821A also shows a PA evolution in a pulse (Sharma et al. 2019). In this work, we find the PD of off-axis bursts will produce a PD turnover during a pulse, which means the PA changes by an angle of 90$^\circ$. This is due to the orientation changes of dominated field (see Fig. 8). Actually at a more later time, the PA will change back again due to the orientation changes of dominated field (see Fig. 8), but the change is not easy to detect due to very low flux and possibly superposed by other pulses. The PA evolution of GRB 170114A and 160821A is roughly consistent with our results. It needs to note that this case is for the off-axis bursts. For an off-axis burst, the flux will tend to be lower, the duration of a pulse will be stretched and the spectrum will be softer due to the relativistically Doppler effect and the propagation effect, assuming the on-axis bursts have the canonical parameters including energy, spectral peak, jet angle, Lorentz factor, etc. It is not clear whether it is the case for the two bursts. But our result suggests that at least a part of bursts (off-axis bursts) should have an evolving PA during a pulse. And those bright and nearby bursts may be detectable. Similar PA evolution with the same reason was also obtained in the optical emission of the reverse shock at the beginning of the GRB afterglow (Lan et al. 2016, 2018). There are some alternative models for PA change, e.g., the orientation change of the MF line due to the magnetic reconnection (Deng et al. 2016). More early, GRB 041219A and GRB 100826 were also found that the PA was changed across the bursts (G\"{o}tz et al. 2009; Yonetoku et al. 2011). But the change happens in different emission episodes, possibly suggesting the MF configuration may be different in the episodes.

The PDs of the five POLAR bursts are generally lower (Zhang et al. 2019a), compared with the observations by other polarimeters, such as AstroSat-CZTI, IKAROS-GAP and INTEGRAL-IBIS. The reason is unknown. It is also unclear whether all GRBs have PD evolution with very small scales. Identifying the evolution requires more sensitive polarimeters and time-resolved polarization with shorter time scales. If such sensitivity can be attained, the theoretically expected rising in the PD curve at the very beginning during a pulse may also be detected. On the other hand, the high-precision time-resolved spectrum with short enough time will also show how the Band spectrum is formed. The combination of the PD and the spectra data will uncover the GRB radiation mechanism and the MF structure. This might be achieved by the next-generation polarimeters such as the enhanced X-ray Timing and Polarimetry mission (eXTP, Zhang et al. 2019b).

\section*{Acknowledgements}
We thank the referee for their valuable comments that improve the paper.
This work was supported by the National Natural Science Foundation of China (No. U1831135), Yunnan Natural Science Foundation (2014FB188).

\clearpage

\begin{deluxetable}{cccccccccccc}
	%\rotate
	\tablecolumns{9}
	\setlength{\tabcolsep}{9pt}
	\tablewidth{0pc}
	\tablecaption{The parameters of the models}
	\tabletypesize{\scriptsize}%footnotesize}
	\tablehead{
		\colhead{Model}                  &
		\colhead{$\Gamma$}                 &
		\colhead{$\gamma_m^\prime$($10^4$)} &
		\colhead{$B_0^\prime$($10^2G$)}&
		%    \colhead{$N_{inj}^\prime$($10^{47}s^{-1}$)}          &
		%    \colhead{$Q_0$($10^{56}s^{-1}$)}      &
		\colhead{a}&
		\colhead{ADI} &
		\colhead{SSC}
	}
	\startdata
	M2(Geng18) & 300 & 10 & 0.3  & 1.0 & Yes & No \\
	M3(Geng18) & 300 & 10 & 0.3  & 1.0 & Yes & Yes \\
	\enddata
	\tablecomments{This group of calculations are just for testing our codes by taking the same parameters as Geng18 (Table 1).
		Note that $R_{s} \neq R_{0}$ is adopted in this group of calculations, where $R_{s}=10^{14}\rm{cm}$ and $R_{0}=10^{15}\rm{cm}$,
		but $R_{s} = R_{0}$ is commonly used in the left part of this paper.}
\end{deluxetable}

\begin{deluxetable}{cccccccccccc}
	%\rotate
	\tablecolumns{9}
	\setlength{\tabcolsep}{9pt}
	\tablewidth{0pc}
	\tablecaption{The parameters of group 1 ($R_{0}=10^{14}cm$)}
	\tabletypesize{\scriptsize}%footnotesize}
	\tablehead{
		\colhead{Model}                  &
		\colhead{$\Gamma$}                 &
		\colhead{$\gamma_m^\prime$($10^4$)} &
		\colhead{$B_0^\prime$($10^2G$)}&
		%    \colhead{$N_{inj}^\prime$($10^{47}s^{-1}$)}          &
		%    \colhead{$Q_0$($10^{56}s^{-1}$)}      &
		\colhead{a}&
		\colhead{ADI} &
		\colhead{SSC}
	}
	\startdata
	M1R14$\Gamma$300(0.0) & 300 & 1 & 20 & 0.0 & No & No \\
	M2R14$\Gamma$300(0.0) & 300 & 1 & 20 & 0.0 & Yes & No \\
	M2R14$\Gamma$300(1.0) & 300 & 1 & 20 & 1.0 & Yes & No \\
	M2R14$\Gamma$300(1.2) & 300 & 1 & 20 & 1.2 & Yes & No \\
	M2R14$\Gamma$300(1.5) & 300 & 1 & 20 & 1.5 & Yes & No \\
	M3R14$\Gamma$300(0.0) & 300 & 2.6 & 3 & 0.0 & Yes & Yes \\
	M3R14$\Gamma$300(1.0) & 300 & 2.6 & 3 & 1.0 & Yes & Yes \\
	M3R14$\Gamma$300(1.2) & 300 & 2.6 & 3 & 1.2 & Yes & Yes \\
	M3R14$\Gamma$300(1.5) & 300 & 2.6 & 3 & 1.5 & Yes & Yes \\
	\enddata
	\tablecomments{The parameters of the models which with initial emission radius $R_{s}=R_{0}=10^{14}\rm{cm}$.}
\end{deluxetable}

\begin{deluxetable}{cccccccc}
	%\rotate
	\tablecolumns{9}
	\setlength{\tabcolsep}{9pt}
	\tablewidth{0pc}
	\tablecaption{The parameters of group 2 ($R_{0}=10^{15}\rm{cm}$)}
	\tabletypesize{\scriptsize}%footnotesize}
	\tablehead{
		\colhead{Model}                  &
		\colhead{$\Gamma$}                 &
		\colhead{$\gamma_m^\prime$($10^4$)} &
		\colhead{$B_0^\prime$($10^2G$)}&
		%    \colhead{$N_{inj}^\prime$($10^{47}s^{-1}$)}          &
		%    \colhead{$Q_0$($10^{56}s^{-1}$)}      &
		\colhead{a}&
		\colhead{ADI} &
		\colhead{SSC}
	}
	\startdata
	M2R15$\Gamma$500(0.0) & 500 & 2 & 3 & 0.0 & Yes & No \\
	M2R15$\Gamma$500(1.0) & 500 & 2 & 3 & 1.0 & Yes & No \\
	M2R15$\Gamma$500(1.2) & 500 & 2 & 3 & 1.2 & Yes & No \\
	M2R15$\Gamma$500(1.5) & 500 & 2 & 3 & 1.5 & Yes & No \\
	M3R15$\Gamma$500(0.0) & 500 & 5 & 0.6 & 0.0 & Yes & Yes \\
	M3R15$\Gamma$500(1.0) & 500 & 5 & 0.6 & 1.0 & Yes & Yes \\
	M3R15$\Gamma$500(1.2) & 500 & 5 & 0.6 & 1.2 & Yes & Yes \\
	M3R15$\Gamma$500(1.5) & 500 & 5 & 0.6 & 1.5 & Yes & Yes \\
	\enddata
	\tablecomments{The parameters of the models which with initial emission radius $R_{s}=R_{0}=10^{15}\rm{cm}$.}
\end{deluxetable}

\clearpage
\begin{figure*}[t!]
	\begin{center}
		\includegraphics[angle=0,width=0.9\textwidth]{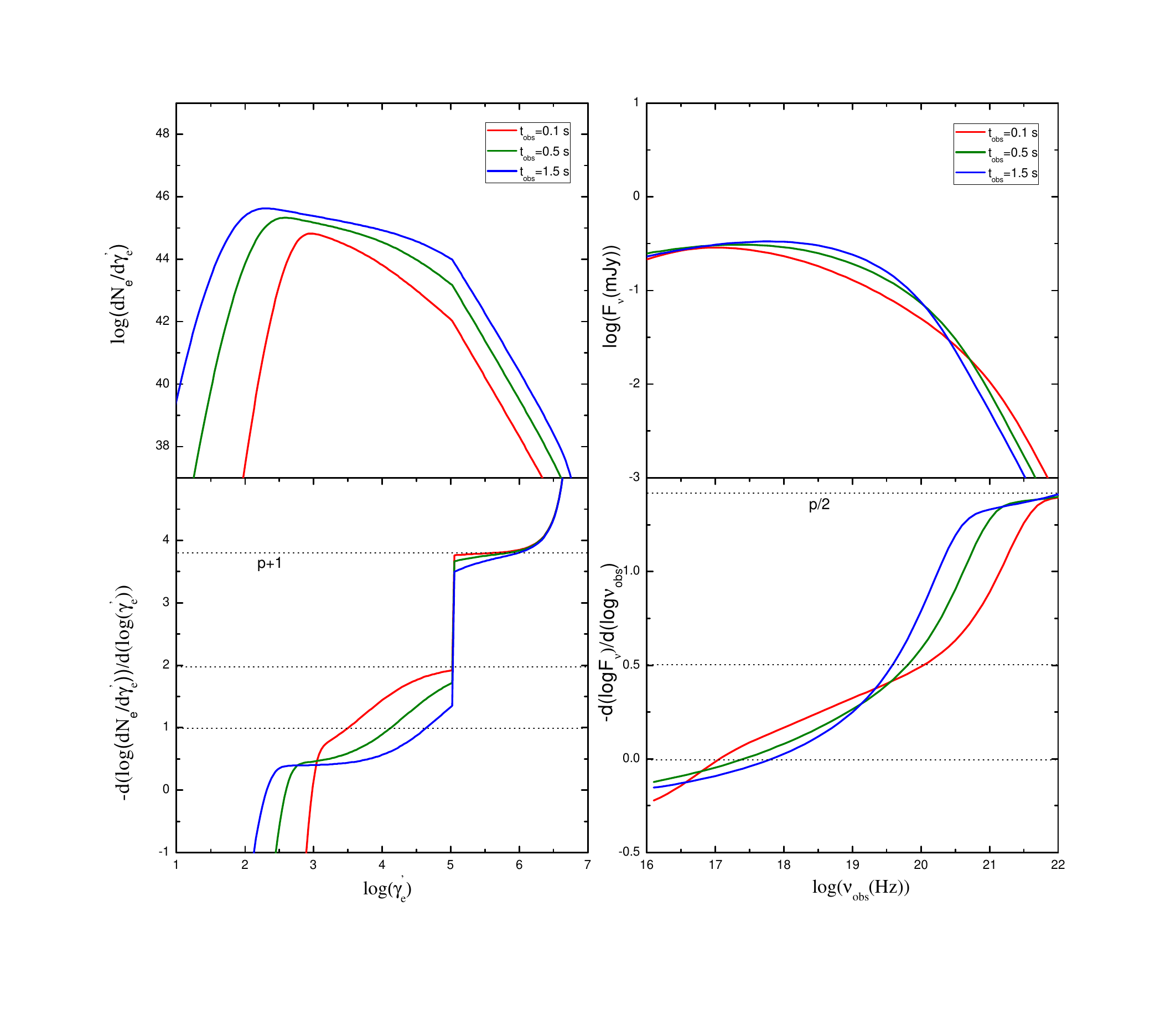}
	\end{center}
	\vglue -0.7cm
	\caption{ Evolution of electron distribution (the upper left panel) and corresponding flux density spectra (the upper right panel) for the model of M2 in Table 1.
		The bottom left and right panels show the negative spectral index of the electron distribution and flux density spectra, respectively. As shown in the figures, our numerical results are consistent with Geng18.}
\end{figure*}

\clearpage
\begin{figure*}[t!]
	\begin{center}
		\includegraphics[angle=0,width=0.9\textwidth]{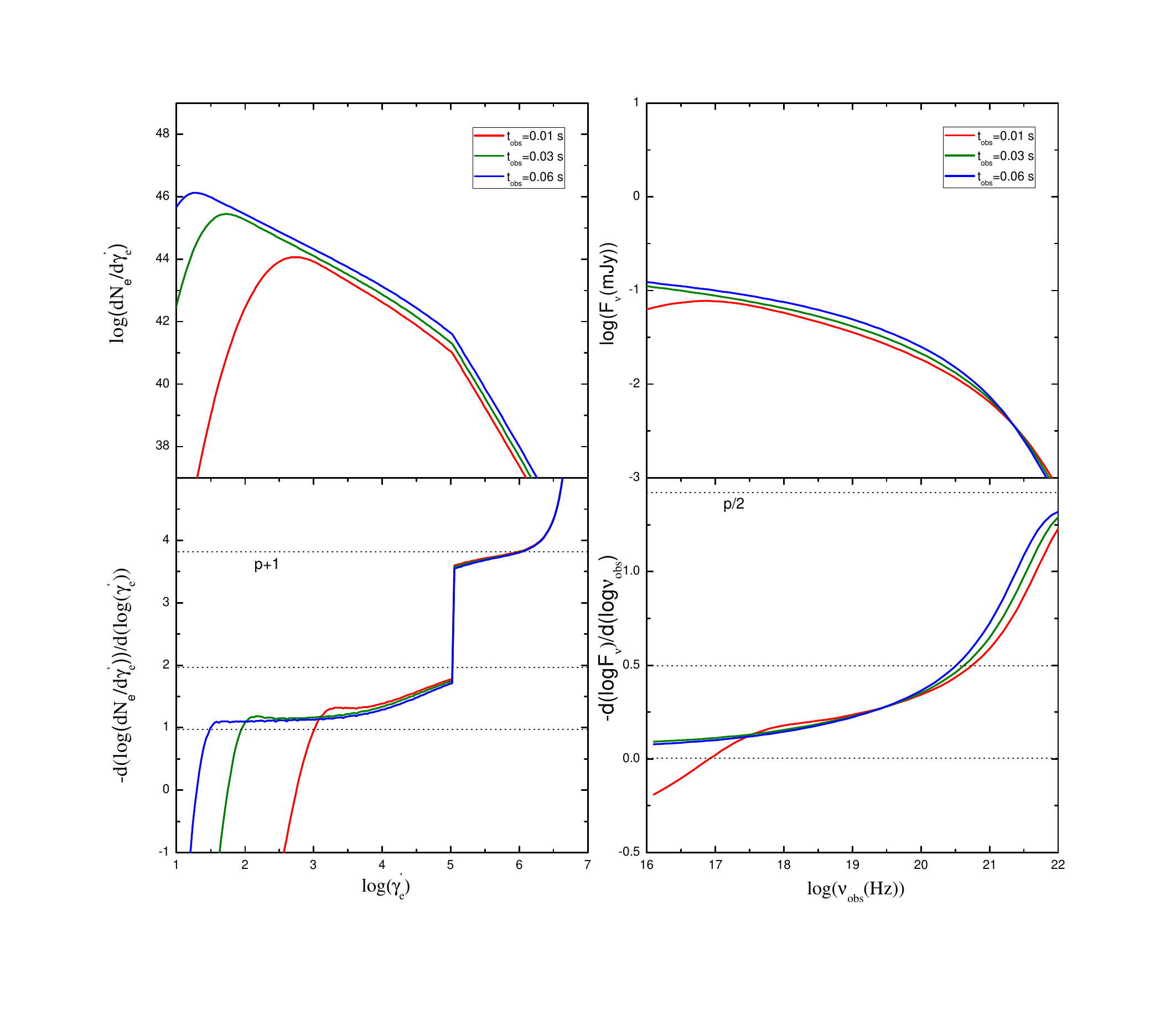}
	\end{center}
	\vglue -0.7cm
	\caption{ Same as Fig. 1, but for the model of M3 in Table 1.}
\end{figure*}

\clearpage
\begin{figure*}[t!]
	\begin{center}
		\includegraphics[angle=0,width=0.9\textwidth]{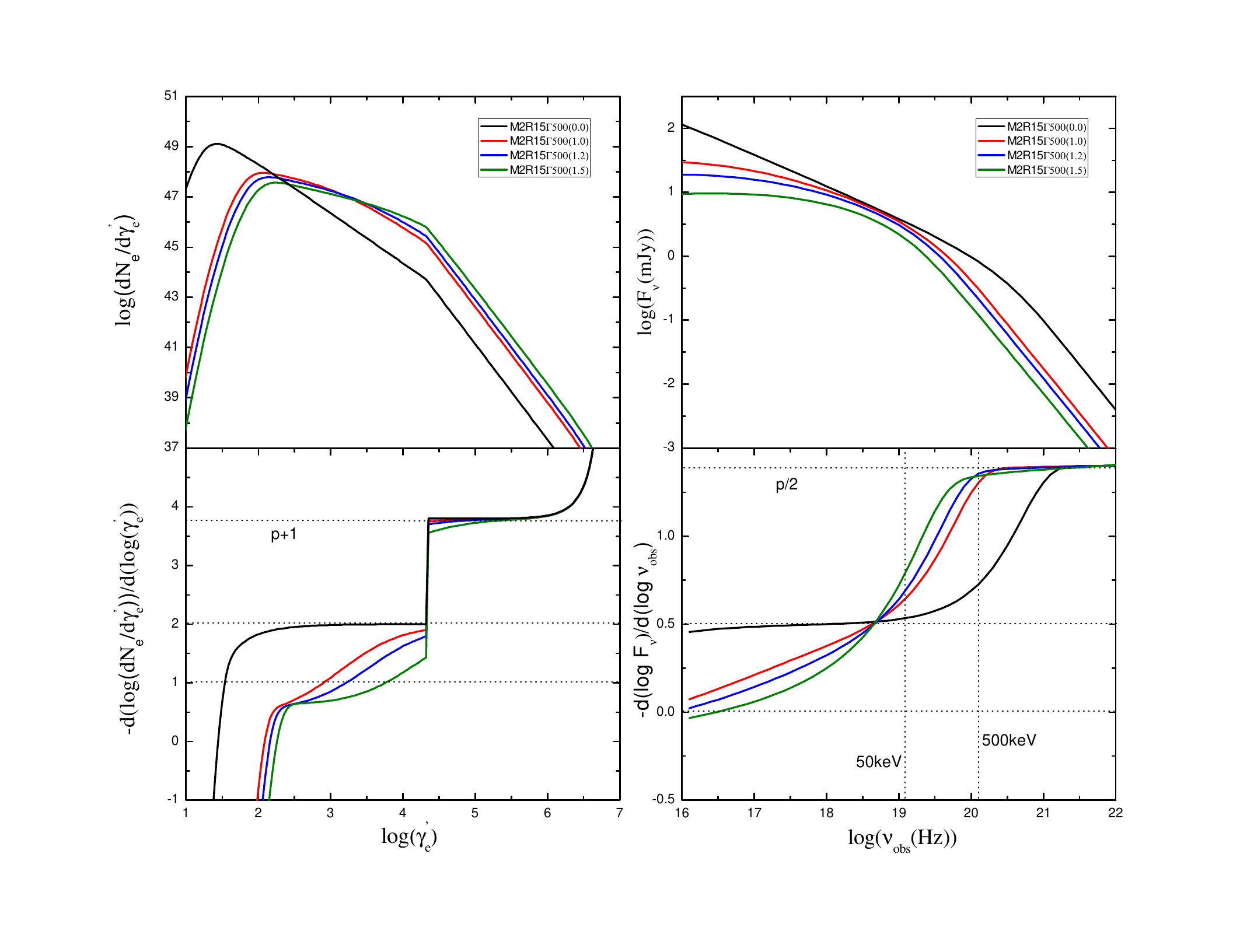}
	\end{center}
	\vglue -0.7cm
	\caption{ The electron distributions and the corresponding flux density spectra for the models M2R15$\Gamma$500(0.0), M2R15$\Gamma$500(1.0), M2R15$\Gamma$500(1.2) and M2R15$\Gamma$500(1.5) at the observed time of $t_{\rm{obs}}=1.0 s$. }
\end{figure*}
%\footnotesize

\clearpage
\begin{figure*}[t!]
	\begin{center}
		\includegraphics[angle=0,width=0.9\textwidth]{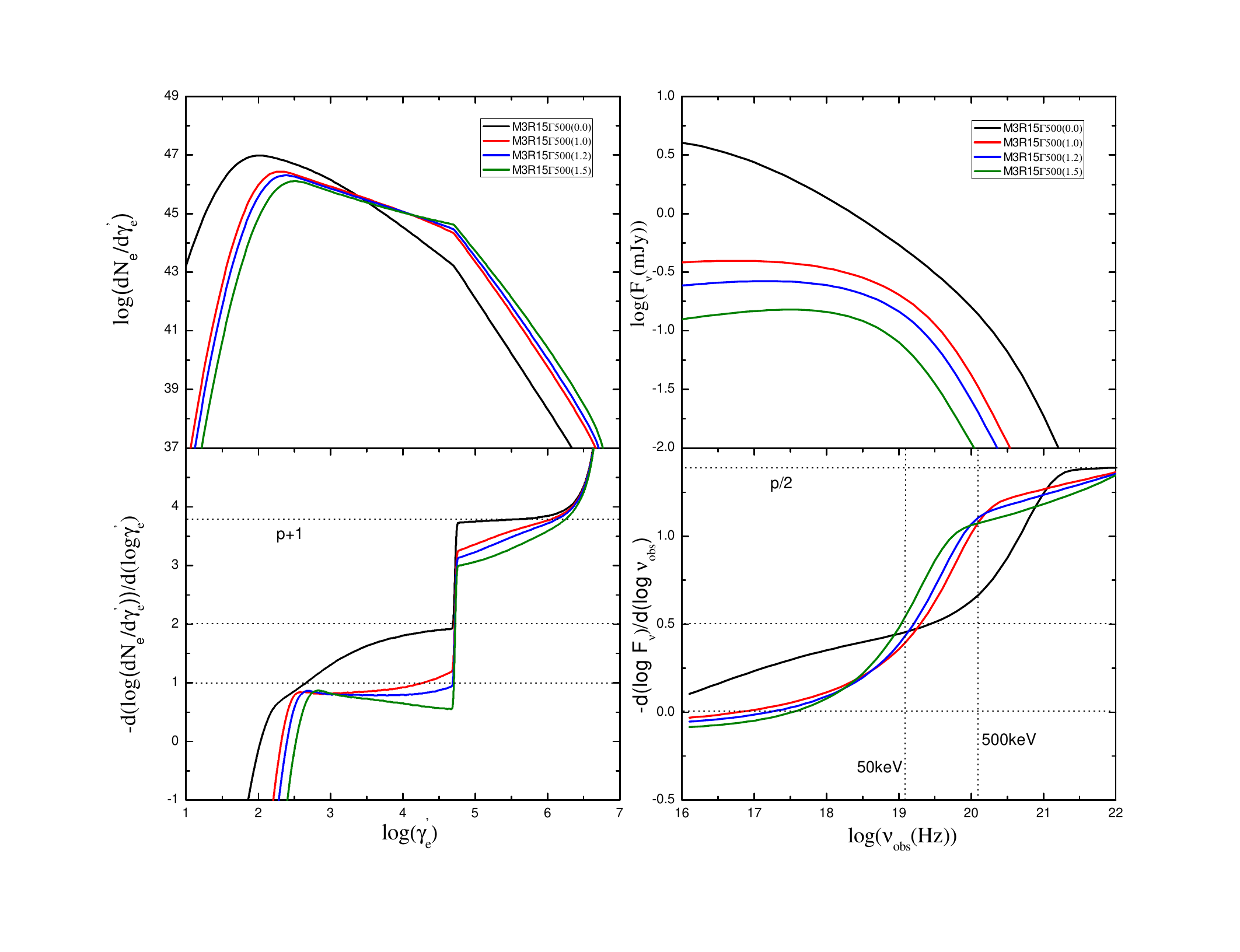}
	\end{center}
	\vglue -0.7cm
	\caption{ Same as Fig. 3, but for the models of M3R15$\Gamma$500(0.0), M3R15$\Gamma$500(1.0), M3R15$\Gamma$500(1.2) and M3R15$\Gamma$500(1.5). }
\end{figure*}

\clearpage
\begin{figure*}[t!]
	\begin{center}
		\includegraphics[angle=0,width=0.9\textwidth]{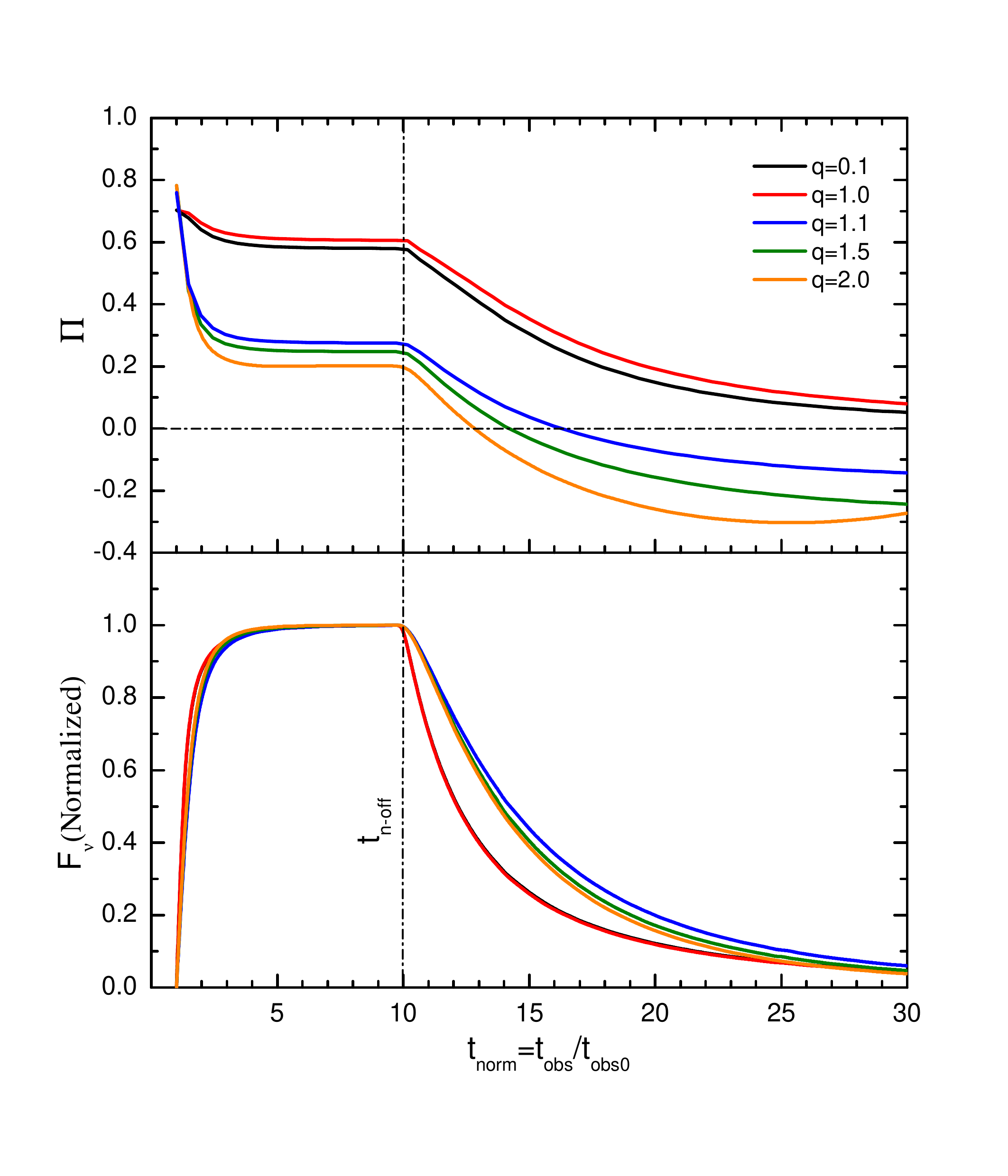}
	\end{center}
	\vglue -0.7cm
	\caption{ Instantaneous PDs (upper panel) and the corresponding normalized (normalized to the peak flux) light curves (bottom panel) for the constant MF model M1R14$\Gamma$300(0.0) (see Table 2.) for different viewing angles (different $q$). Here we take $t_{\rm{n-off}}=t_{\rm{off}}/t_{\rm{obs0
	}}=R_{\rm{off}}/R_{\rm{s}}=10$. See the context for the definition of $t_{\rm{off}}$ and $t_{\rm{obs0}}$. As shown in the figure, the instantaneous PD in the off-axis case ($q>1$) will evolve from a positive value to a negative one, suggesting that the PA changes by an angle of $90^\circ$.}
\end{figure*}

\clearpage
\begin{figure*}[t!]
	\begin{center}
		\includegraphics[angle=0,width=0.9\textwidth]{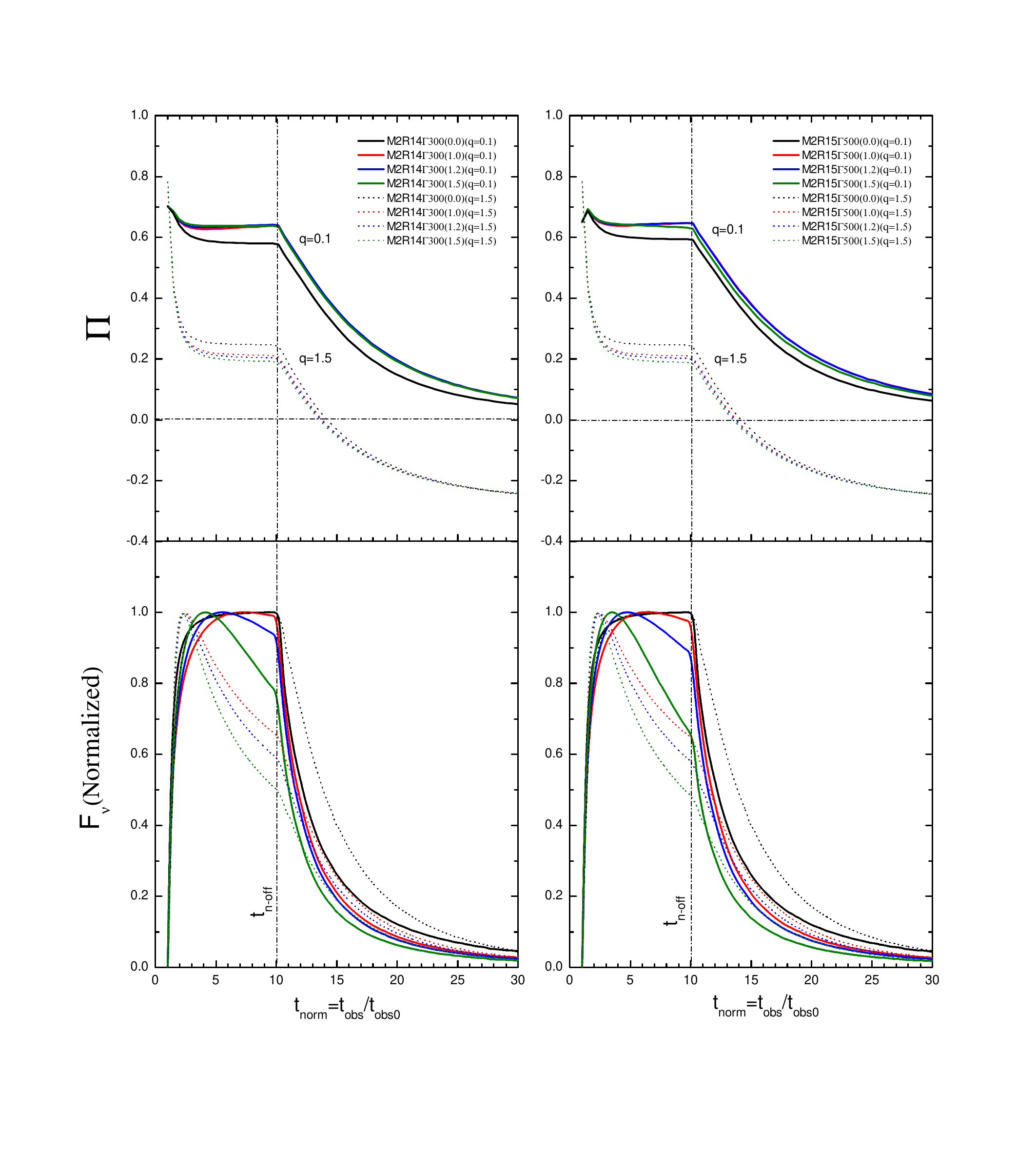}
	\end{center}
	\vglue -0.7cm
	\caption{ Same as Fig. 5, but for model M2 (see Tables 2 and 3). The left panels: model M2 with the initial emission radius of $R_0=10^{14}$ cm. The right panels: model M2 with $R_0=10^{15}$ cm. As shown in the figure, the PDs for the decaying MF models are a little higher than the constant MF model for the on-axis case($q=0.1$). On the contrary, the PDs for the decaying MF models are a little lower than the constant MF model for the off-axis case ($q=1.5$).}
\end{figure*}

\clearpage
\begin{figure*}[t!]
	\begin{center}
		\includegraphics[angle=0,width=0.9\textwidth]{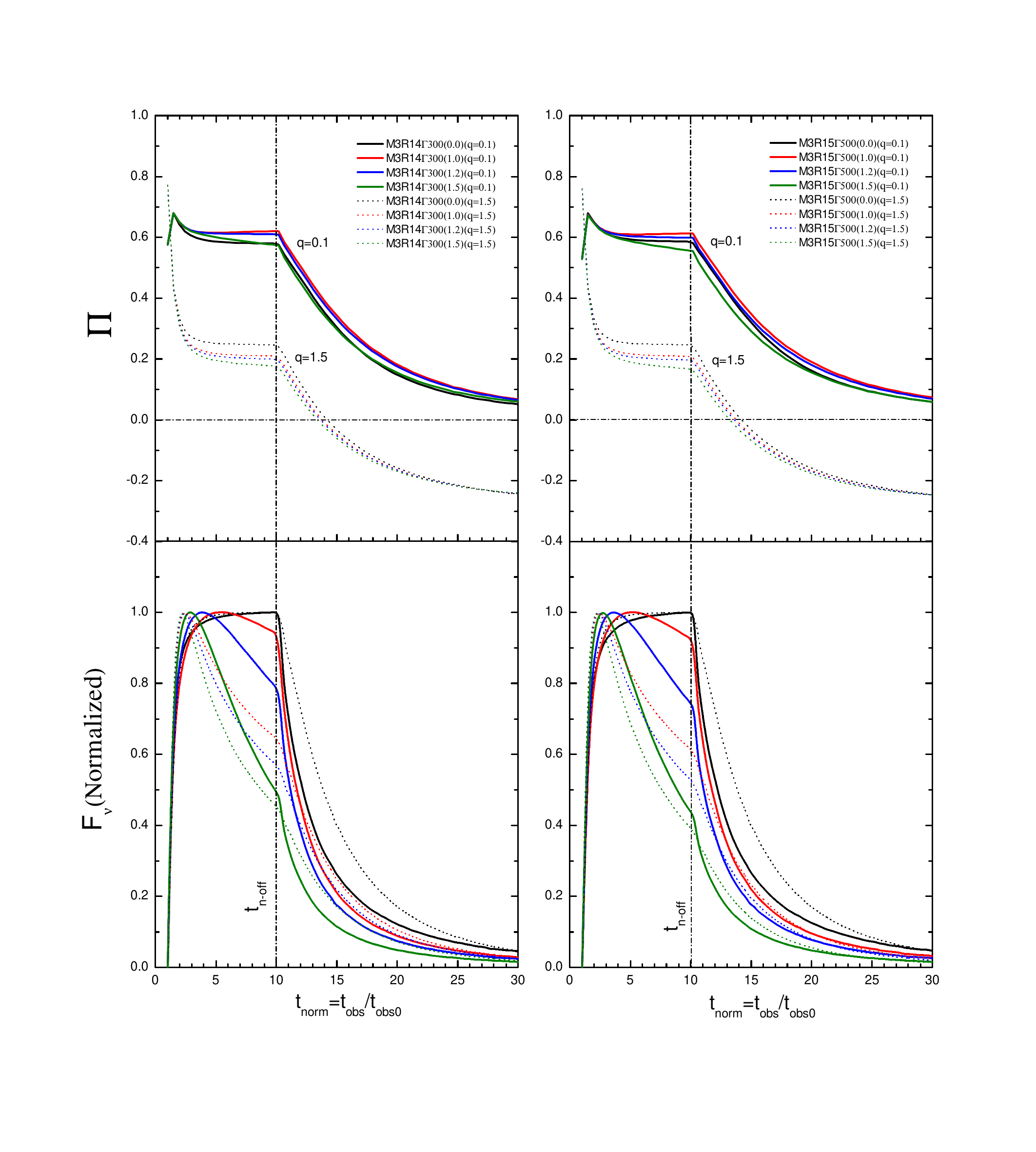}
	\end{center}
	\vglue -0.7cm
	\caption{Same as Fig. 6, but for the model M3.}
\end{figure*}

\clearpage
\begin{figure*}[t!]
	\begin{center}
		\includegraphics[angle=0,width=0.9\textwidth]{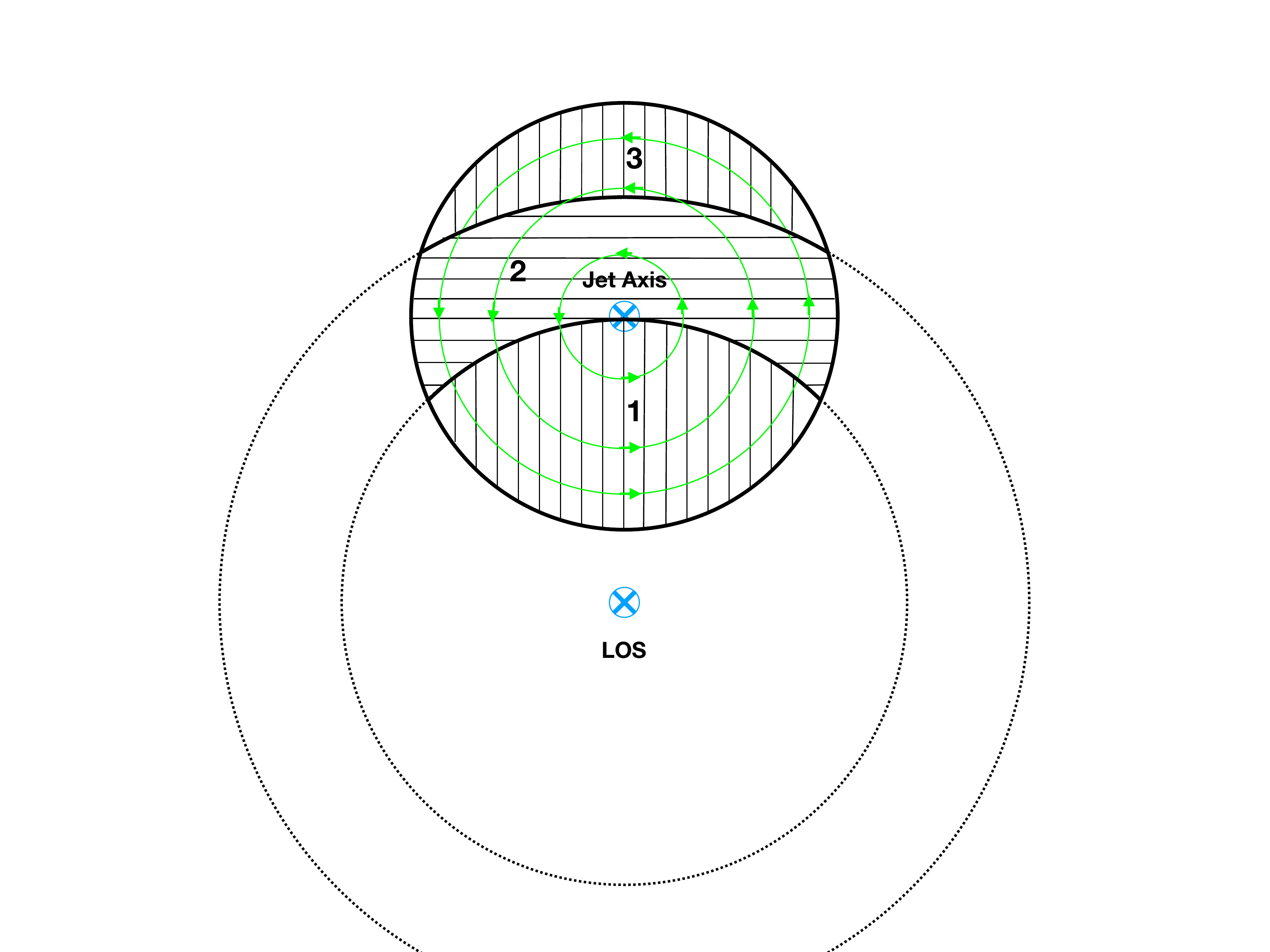}
	\end{center}
	\vglue -0.7cm
	\caption{ Polarization schematic diagram for the off-axis case. The green circles stand for the toroidal MF. The hatched regions with horizontal and vertical lines represent the polarization vectors.
	The polarization region is divided into three areas (1,2,3). The photons from the three areas come into our sight in sequence. The polarization vectors vary
	from the vertical direction to the horizontal direction, and then back to the vertical direction. The corresponding PD varies from positive to negative and then back to positive again at a very late time. }
\end{figure*}

\clearpage
\begin{figure*}[t!]
	\begin{center}
		\includegraphics[angle=0,width=0.9\textwidth]{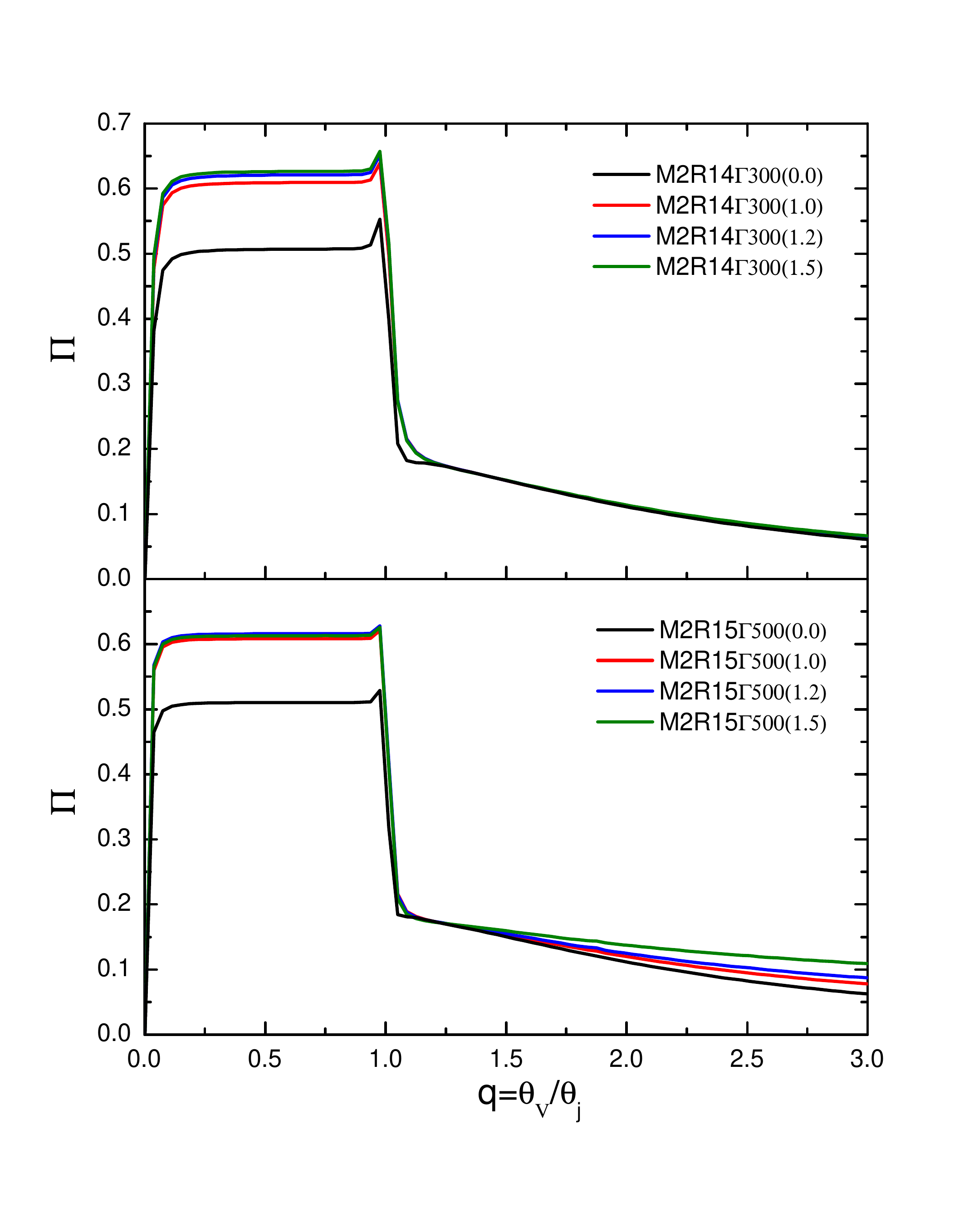}
	\end{center}
	\vglue -0.7cm
	\caption{ Time-averaged PDs with different viewing angles for the model M2. The upper panel: model M2 with the initial emission radius of $R_0=10^{14}$ cm. The bottom panel: model M2 with $R_0=10^{15}$ cm.
	As is shown, the time-averaged on-axis PDs are $\sim 0.6$ for the decaying MF models and $\sim 0.5$ for the constant MF model, while the off-axis PDs sharply decrease down to $\sim 0.1$.}
\end{figure*}

\clearpage
\begin{figure*}[t!]
	\begin{center}
		\includegraphics[angle=0,width=0.9\textwidth]{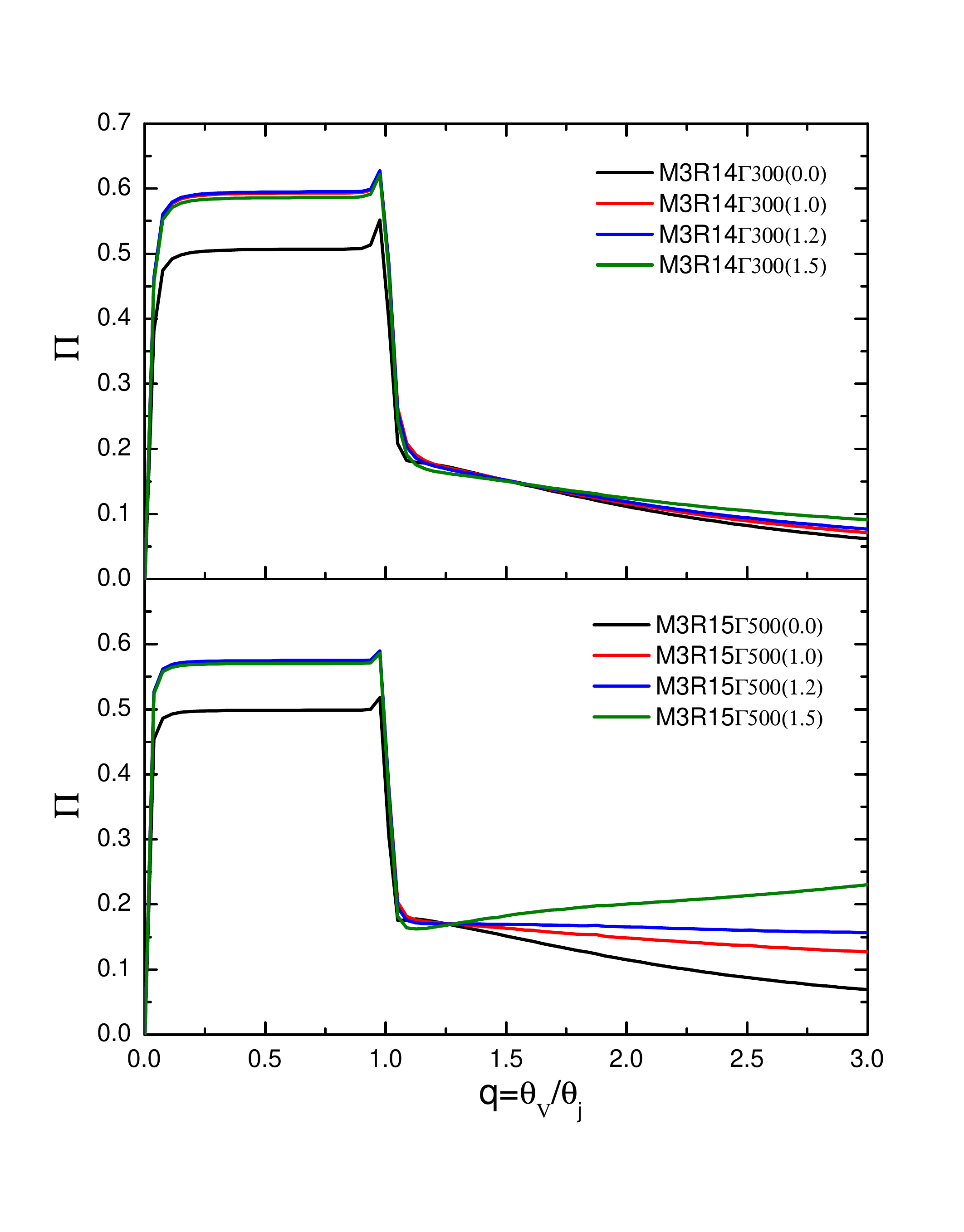}
	\end{center}
	\vglue -0.7cm
	\caption{ Same as Fig. 9, but for the model M3.}
	
\end{figure*}

\end{document}